\documentclass[aps,pra,twocolumn,groupedaddress, showpacs]{revtex4}

 \usepackage{graphicx}
 \usepackage{amssymb}
 \usepackage{epstopdf}
\begin{document}

\preprint{LBNL-57137}
\title{Focusing a fountain of neutral cesium atoms with an electrostatic lens triplet}


 \author{Juris G. Kalnins}
 \email{JGKalnins@lbl.gov}
\author{Jason M. Amini}
 \email{JAMaddi@lbl.gov}
 \altaffiliation[Also at ]{Physics Department, University California at Berkeley, Berkeley, CA 94720.}
\author{Harvey Gould}
 \email{HAGould@lbl.gov}
\affiliation{Mail stop 71-259, Lawrence Berkeley National Laboratory,  Berkeley, CA 94720}

%

\date{\today}

\begin{abstract}
An electrostatic lens with three focusing elements in an alternating-gradient configuration is used to 
focus a fountain of cesium atoms in their ground (strong-field-seeking) state. The lens electrodes are shaped to produce only sextupole plus dipole equipotentials which avoids adding the unnecessary nonlinear forces present in cylindrical lenses. Defocusing between lenses is greatly reduced by having all of the main electric fields point in the same direction and be of nearly equal magnitude. The addition of the third lens gave us better control of the focusing strength in the two transverse planes and allowed focusing of the beam to half the image size in both planes. 
The beam envelope was calculated for lens voltages selected to produced specific focusing properties. The calculations, starting from first principles, were compared with measured beam sizes and found to be in good agreement. Application to fountain experiments, atomic clocks, and focusing polar molecules in strong-field-seeking states is discussed.
\end{abstract}


\maketitle
\section{\label{Intro}Introduction}
The force exerted by a static electric field gradient on a beam of slow neutral atoms can produce useful focusing\cite{noh00}. This force arises from the interaction of the electric field gradient with the induced electric dipole moment of the atom. Because no external magnetic fields are introduced, electric field gradient focusing is suitable for magnetic-field sensitive experiments such as atomic fountain clocks and electron electric dipole moment experiments.

In cesium fountain atomic clocks, focusing could be used to reduce the cesium-cesium scattering. Cesium-cesium scattering produces a density-dependent frequency shift\cite{gibble93,ghezali96,leo01} and has been listed as the largest contribution to the uncertainty in the NIST-F1 fountain clock\cite{jefferts02}. Although all of the cesium atoms that enter the clock interaction region contribute to the cesium-cesium scattering frequency shift, only the small fraction of the atoms return to exit the interaction region and contribute to the signal. In NIST-F1 about 90 percent of the atoms have been reported lost \cite{jefferts02}. If the non-returning atoms are removed before they enter the interaction region, the atoms would not contribute to the cesium-cesium scattering and the systematic would be reduced.

The unwanted atoms could be removed by having a set of electrostatic lenses expand the cesium fountain beam, thereby making it into a nearly parallel beam, and then using a collimator to remove all but the center portion of the beam.
The atoms that contribute to the signal would be in the center of the beam and pass through the collimator.  Section \ref{Measurements} and especially Fig.\ref{tunec10} show examples of focusing to produce nearly parallel beams of cesium atoms (in one dimension).  

Controlling the beam profile by focusing lenses and collimator will also greatly reduce the dependence of the beam trajectory on variations in source position, source size, and initial beam divergence. Such variations will change the number of atoms that pass through the interaction region, but will have little influence on the beam trajectory. With more stable beam trajectories it may be possible to boost the beam intensity by using a trap as the atom source, rather than a molasses.


Fountain experiments to search for an electron electric dipole moment or to measure polarizability\cite{amini03} require a region of strong electric field. The atoms defocus as they enter and exit the field\cite{maddi99}, robbing the experiment of needed intensity.  A set of lens elements can largely compensate for the defocusing, maintaining beam intensity. 

Electric field gradient focusing is also useful in fountains of alkali atoms containing more than one magnetic substate ($m_F$). The forces produced, by an electric field gradient, on alkali atom and other $J \le 1/2$ ground state atoms are essentially independent of the magnetic quantum number, $m_F$\cite{angel68}. Magnetic field gradients produce forces that are strongly dependent on the magnetic substate and thus focus each $m_F$ state differently. 

For polar molecules, electric field gradient focusing is the standard method for focusing molecular beams, including thermal beams and jet-source beams. Polar molecules typically have large electric dipole moments and very small magnetic dipole moments The focusing lens triplet described in this paper is suitable for focusing polar molecules in strong-field-seeking states and also polar molecules in some weak-field-seeking states. Molecules such as chlorobenzene and cesium fluoride have small rotational constants and large electric dipole moments. This makes all of their low-lying rotational states strong-field seeking in even a modest electric field. Efficient focusing of these molecules should be possible with our lens design.

In the remainder of the paper we discuss alternating gradient focusing (Section \ref{electrostatic}),  lens design and the choice of electric multipole moments for linear focusing of atoms and 
molecules (Section \ref{design}), beam trajectory calculations (Section \ref{trajectory}) and finally, in Section \ref{experiment},  the construction and operation of a triplet lens in a cesium fountain and comparison between calculation and measurement for different focusing conditions. 
\section{Alternating Gradient Focusing\label{electrostatic}}
The force on the atom in an electric field gradient is:
 \begin{displaymath} \label{1}
{\bf F} = - \nabla W = - \frac{dW}{dE}\nabla E=\alpha E \nabla E\\
\end{displaymath}
Where
 \begin{displaymath}W = -\alpha \frac{E^2}{2}
 \end{displaymath}
is the potential energy of a ground-state atom in an electric field of magnitude 
 $E = (E_x^2 + E_y^2 +E_z^2)^{1/2}$ and 
 $\alpha = (6.611 \pm 0.009)\times 10^{-39}$ Jm$^2$V$^{-2}$ 
 is the cesium static dipole polarizability\cite{amini03}.
 
The force from the electric field gradient on ground state atoms is towards the stronger field (strong-field seeking). To focus ground state atoms, and molecules in strong-field-seeking states, in both transverse directions, it is necessary to use at least two lenses in an alternating gradient configuration\cite{noh00, auerbach66, kakati67, kakati69,gunther72, lubbert75, lubbert78, reuss88, bethlem02a}. Each lens focuses in one transverse direction while defocusing in the other. A single electrostatic lens can not focus  strong field seeking states because an electric field that is uniform along the beam direction cannot also have a \emph{maximum} in both transverse directions. (For atoms and molecules in weak-field-seeking states, a single lens can focus in both transverse directions because an electric field that is uniform along the beam direction can have a minimum in both transverse directions.)

A net focusing in both transverse directions is achieved using alternating gradients so that the focusing (defocusing) in the first lens is in the same direction as the defocusing (focusing) in the second lens.  Atoms defocused, in say the $x$ direction, in the first lens, are further from the centerline in the second lens and experience a stronger restoring force. Atoms focused in the $y$ direction, in the first lens, are closer to the centerline and experience smaller defocusing in the second lens. For best results the forces should be linear in the displacement in $|x|$ and $|y|$ from the $z$ axis.
 
\begin{figure}[b]
\includegraphics [scale = 0.666] {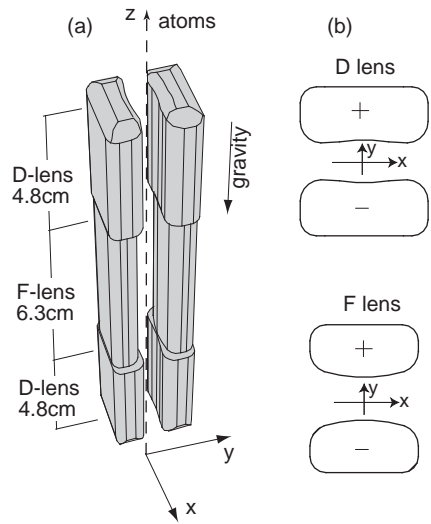}
\caption{\label{allplates}
Drawing of (a) the lens triplet and its coordinate system and (b) the cross-sections for the D lens and F lens electrodes. In  both the D and F lenses, the full gap between the electrodes at $x = 0$ is 8 mm. For atoms and molecules in strong-field-seeking states, the forces in the D lens are towards $y = 0$ and away from $x = 0$ and in the F lens, the forces are away from $y = 0$ and towards $x = 0$.}
\end{figure}

 It is important to note that the force does not depend on the direction of the electric field.
Therefore, all lenses can have their electric fields pointing 
in the same direction as shown in Fig.\ref{allplates}a.
%
%

The focusing direction is determined by the curvature of the electrodes, as shown in Fig.\ref{allplates}b. The F lens focuses atoms and molecules in strong-field-seeking states in the $x$ direction and defocuses them in the $y$ direction, while the D lens defocuses in the $x$ direction and focuses in the $y$ direction. The shapes of the electrodes are carefully chosen to produce forces on the atoms that have the smallest nonlinearities in both $x$ and $y$. 

Although an F-D or D-F lens doublet can focus a beam in both transverse planes, the ratio of image size to object size (magnification) in the two planes are unequal: if one magnification is less than one, the other will be greater than one. We therefore chose a triplet (D-F-D) lens, shown in Fig.\ref{allplates}, which, with sufficient strength, allows us to focus the beam equally in both planes down to a magnification of 0.5.

\section{\label{design} Lens Design}
To specify the transverse electric fields for the lenses, we expand the electric field potential in multipoles. For an atom or a molecule in a strong-field-seeking state, the electric field that comes closest to producing forces that are linear in displacement from the $z$ axis of the lens is the field generated by the combination of a sextupole potential plus a dipole potential\cite{kalnins02}. (A sextupole alone will defocus in both transverse directions and a quadrupole component is not used because it will deflect the atoms.) A small decapole potential may improve the linearity slightly, but otherwise the presence of higher order terms (found in cylindrical electrodes) produces forces that are non-linear in the displacement. Non-linear forces cause beam loss and larger, more diffuse, beams.

The electric field of a dipole plus sextupole potential is, to lowest order in $x$ and $y$\cite{kalnins02}: \begin{equation} \label{E}
E^2 = E_{y_0}^2 \left[ 1+ 2A_3 ( x^2 - y^2 )\right] + (E_{y_0}^{\prime 2} -E_{y_0} E_{y_0}^{\prime \prime}) y^2
\end{equation}
where the primes denote derivatives in $z$, 
$E_{y_0} (z) \equiv E_y (x=0, y=0, z)$ is the electric field in the $y$ direction along the $z$ axis 
$(x = y = 0)$, and $A_3$ is the sextupole strength relative to the dipole strength. $A_3$ is  positive for a D lens and a negative for an F lens. The last term in Eq.\ref{E} accounts for the defocusing in the $y$ direction, due to the fringe fields, as the atoms enter or exit a lens. 

The transverse shape of the lens electrodes with the dipole plus sextupole potential (Fig. \ref{allplates}) is given by the parameterized curve $(x_p, y_p)$:
\begin{displaymath}
x_p  =  \sqrt{\frac{r_p-y_p-\frac{1}{3}A_3(r_p^3 -y_p^3)}{A_3 y_p}}
\end{displaymath}
where $r_p$ is the half-gap spacing of the electrodes at $x_p =0$.
This gives the top right section of the electrode with the restriction 
$y_p \ge r_p$ for a convex F lens and $r_p \ge y_p > 0$ for a concave D lens.

In our electrodes the half-gap $r_p$ is 4 mm and the sextupole strengths 
are $A_3 = -4.031 \times 10^3$ m$^{-2}$ for the F lens and 
$A_3 = 4.444 \times 10^3$ m$^{-2}$ for the D lens.

Polar molecules in strong-field-seeking states such as J = 0 also require alternating gradient focusing and this has been the subject of a number of 
experiments\cite{auerbach66, kakati67, kakati69,
gunther72, lubbert75, lubbert78, reuss88, bethlem02a}.
Although the potential energy $W$ and hence $dW/ dE$ of polar molecules is of a different form than for atoms, the same lenses can be used for both atoms and polar molecules with only a small difference in linearity\cite{kalnins02}. 
The lens electrodes used in the measurements, described in Section \ref{Measurements}, were originally designed for focusing the $J = 0$ rotational state of a jet-source methyl fluoride beam.
\section{Trajectory Calculations\label{trajectory}}
The Hamiltonian of an atom in the fountain, including the potential due to the electric field in the coordinate system of Fig.\ref{allplates}, is: 
\begin{equation} \label{H}
H = \frac {m |\textbf{v}|^2}{2}  + mgz - \frac {\alpha E^2}{2}
\end{equation}
where $\textbf v$ is the velocity, $m = 2.207 \times 10^{-25}$ kg is the cesium mass and
$g = 9.8$ ms$^{-2}$ is the acceleration of gravity.

The equation of motion for an atom in the fountain is then:
\begin{displaymath}
m \frac{d\textbf v} {dt} = -\nabla H.
\end{displaymath}
Substituting Eq.'s ~\ref{E}, \ref{H} into the equations of motion , the transverse components of the acceleration are:
\begin{eqnarray} \label{motion}
\frac{d v_x}{d t}& = & \frac{\alpha}{m} \left[ 2A_3 E_{y_0}^2 \right]  x \nonumber \\
\frac{d v_y}{d t}& = & \frac{\alpha}{m} \left[ -2A_3 E_{y_0}^2 +(E_{y_0}^{\prime 2}
- E_{y_0} E_{y_0}^{\prime \prime}) \right] y.
\end{eqnarray}

From the conservation of the Hamiltonian in the case of static fields, we obtain the central longitudinal velocity in the fountain:
\begin{displaymath} \label{longitudinal}
v_z^2(z) = v_{z_0}^2 -2g z + \frac{\alpha}{m} E_{y_0}^2(z)
\end{displaymath}
where $v_{z_0}$ is a constant ($v_z = v_{z_0}$ at $z = 0$ and $E_{y_0}(0) =0$).

We now obtain the equations of motion for the beam envelope of a monoenergic beam distribution in  4-dimensional phase space $(x, v_x, y, v_y)$.
Taking the second moments of the beam distribution
$\sigma_x^2(z) \equiv <x^2>$, $\sigma_{xv_x}(z) \equiv <xv_x>$, $\sigma_{v_x}^2(z) \equiv <v_x^2>$, etc., and from  Eq.(\ref{motion}), we obtain the root-mean-square (RMS) envelope equations\cite{sacherer71} in the $x$ direction:
\begin{eqnarray} \label{envelope}
v_z(\sigma_x^2)^\prime - 2\sigma_{xv_x} & = & 0 \nonumber \\
v_z(\sigma_{xv_x})^\prime - \sigma_{v_x}^2 + G_x(z)\sigma_x^2 & = &0 \nonumber \\
v_z(\sigma_{v_x}^2)^\prime + 2G_x(z)\sigma_{xv_x} & = & 0
\end{eqnarray}
 where the primes denote derivates in $z$. Similar equations hold for the $y$ direction.
 
 The linear focusing fields $G_x(z)$ and $G_y(z)$ are given by:
\begin{eqnarray} \label{correction}
G_x(z) & = & - \frac{\alpha}{m} 2 a_x  A_3 E_{y_0}^2  \nonumber \\
G_y(z) & = &  \frac{\alpha}{m} \left[ 2 a_yA_3 E_{y_0}^2 - \left(E_{y_0}^{\prime 2} - E_{y_0}  E_{y_0}^{\prime \prime} \right)\right]
\end{eqnarray}
where we have added the correction terms $(a_x, a_y)$, which are different from one only if the actual lens field deviates significantly from the pure dipole-plus sextupole field (for which $a_x = a_y$ = 1).

The three moments in the $x$ direction define the RMS emittance:
$\epsilon_{xv_x} = \sqrt{\sigma_x^2 \sigma_{v_x}^2 -\sigma_{xv_x}^2} $ 
which, from Eq. \ref{envelope}, is conserved, i.e., $\epsilon ^{\prime}_{xv_x} = 0$.
This means that for linear forces, the phase space density and 
emittance area (e.g., $\epsilon_{xv_x} = \sigma_x \sigma_{v_y}$ 
at a focus where $\sigma_{xv_x} = 0$)
remain constant (and similarly for the $y$ direction).

In our experiment, the cesium atoms, initially launched into the fountain as a small bunch, with a non-zero temperature, become distributed in longitudinal position and velocity $(z, v_z)$. This will smear out the transverse edges of the beam because the low velocity atoms will be (slightly) over focused and the high velocity atoms will be (slightly) underfocused.

\section{Experiment\label{experiment}}
\subsection{\label{Apparatus} Apparatus}
\begin{figure}[h]
\includegraphics [scale = 0.8] {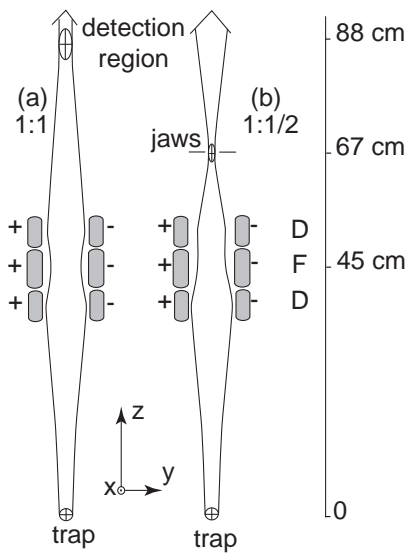}
\caption{\label{apparatus}
Schematic of the fountain apparatus. The cesium atoms are launched at 4.7 m/s and pass through a square aperture (not shown) to remove those atoms whose trajectory falls outside the aperture of the triplet lens situated a short distance above. In (a) the magnification is one and the cesium atom profile is measured by scanning the laser beam near the focus. In (b) the magnification is 0.5 and, because the cesium atoms focus below the laser beam, their profile is determined by scanning a jaw separately in the $x$ and $y$ directions at this focus and measuring the transmitted beam with the laser probe.}
\end{figure}

We measured the focusing properties of the triplet of lenses, shown in Fig.~\ref{allplates}, by passing bunches of slow, cold, cesium atoms through the lenses and into a detection region as shown and described in Fig.~\ref{apparatus}. Our basic cesium atom fountain, which uses a vapor-capture magneto-optic trap and launches the atoms as a moving molasses, has been previously described in Ref.'s \cite{maddi99, amini03}.

%
%
The order of lenses in the triplet can be either D--F--D or F--D--F. We chose D--F--D because it makes the beam smallest in the $y$ direction inside the lenses, where the electrodes limit the aperture.

The fringe field defocusing (in the $y$ direction) is significant, especially when the beam size is a large fraction of the gap spacing, and/or the beam velocity is low. Therefore it is important to minimize the size and number of fringe fields and to accurately include the remaining defocusing effects in the calculations.

To minimize $y$ direction defocusing between the F and D lenses, the changes in the electric field between lenses should be kept small\cite{maddi99}. We therefore used the same orientation of electric fields in all three lenses and designed the lenses to operate at nearly the same voltages. 
As a result of using similar lens voltages, the spacing between F and D lenses could be reduced to one millimeter and the electrode ends where the F and D lenses faced each other needed very little rounding (about one millimeter radius). 

With only a small change in the magnitude of the electric field between lenses and with all of the main (dipole) fields in the same direction, the atoms traveling between the F and D lenses experience only a very small fringe-field-induced defocusing  (the second term in Eq.~\ref{correction}). To minimize the $y$ direction fringe-field defocusing at the entrance and exit of the triplet, the outer ends of the D lenses were rounded to a radius of four millimeters.

In the D lens, the electrodes are concave and had to be terminated at some reasonable minimum gap between the electrodes (Fig.\ref{allplates}). We mistakenly chose a simple rounding of the transverse edges of the D lens. This perturbed the central field, reducing the linear focusing strengths $a_x$ and $a_y$ in Eq.~\ref{correction} from their ideal value of 1 to as low as 0.8. These reduced values of $a_x$ and $a_y$ were used in the trajectory calculations. 

The simple rounding also introduced their own nonlinearities in the forces.  Further modeling has shown that a small correction to the rounding of the transverse edges can restore focusing strength and eliminate these nonlinearities. However, our existing lenses could not be easily modified to include this correction, so instead the trajectory calculations used the reduced focusing strength and increased nonlinearities.
%
\begin{figure} [b]
\includegraphics [scale =0.75]  {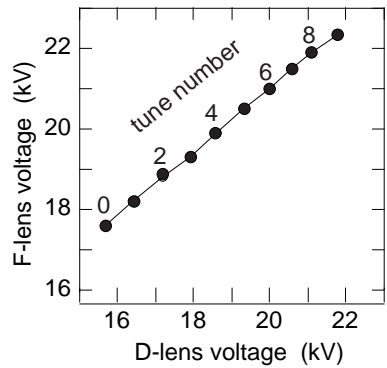}
\caption{\label{tune}
The series of voltages applied to the F and D lenses for a double focus. Each combination of voltages that generates a focus in both $x$ and $y$ at the same vertical location is a tune. Increasing tune number corresponds to higher voltages and increased focusing.  Tune number 6 gives a double waist just ahead of the laser probes and with a magnification of one.}
\end{figure}
%

The triplet electrodes are made from an aluminum alloy whose electric field holding property is adequate for our $\le$ 5.5 MV/m electric fields. Aluminum alloy is less expensive to machine than stainless steel and is nonmagnetic. The electrodes are supported by ceramic spacers attached to a grounded frame. Both of the D lenses are connected to the same pair of positive and negative high voltage supplies (largely motivated by a limited access for high voltage connections in our fountain apparatus). The F lens, between the D lenses, has a separate pair of  high voltage power supplies to allow for tuning the triplet.  Each lens has an 8 mm spacing between electrodes ($r_p = $ 4 mm) along the centerline.  The lens aperture and overall length is smaller than ideal, having been restricted by the limited space in our existing fountain apparatus. This limits the intensity but does not change the focusing principles.
\subsection{\label{Measurements} Measurements}
We performed three sets of focusing measurements. In the first set, we imaged the cesium trap with a magnification of about one in both $x$ and $y$. In the second set we raised the voltages, thereby increasing the focusing strength and imaged the trap with a magnification of about 0.5. In the third set, we adjusted the voltages to produce: first, a parallel beam in the $y$ direction with a waist in the $x$ direction, then a nearly parallel beam in the $x$ direction with a waist in the $y$ direction.
%
\begin{figure} [b]
\includegraphics [scale =0.4] {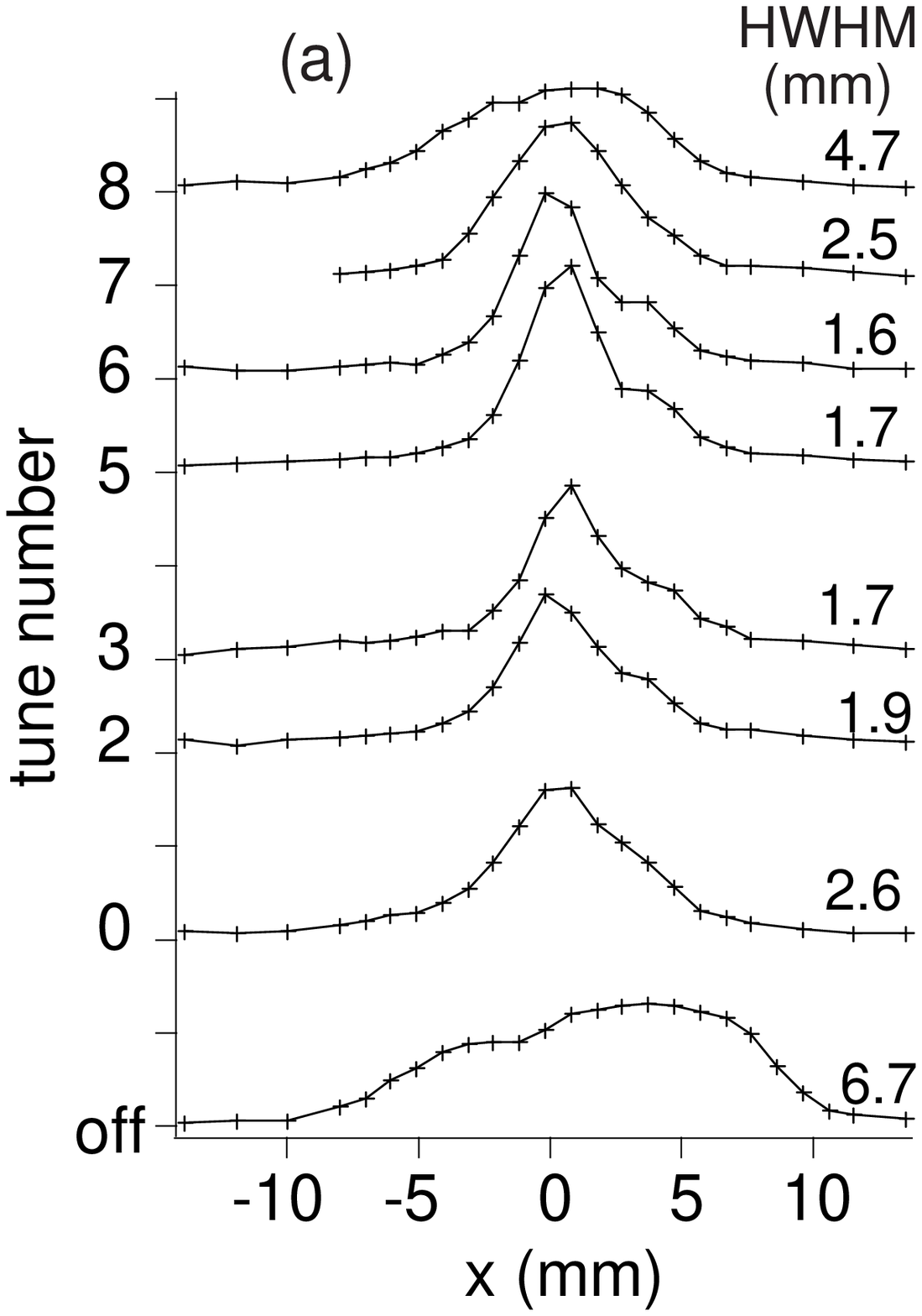}
\includegraphics [scale =0.4] {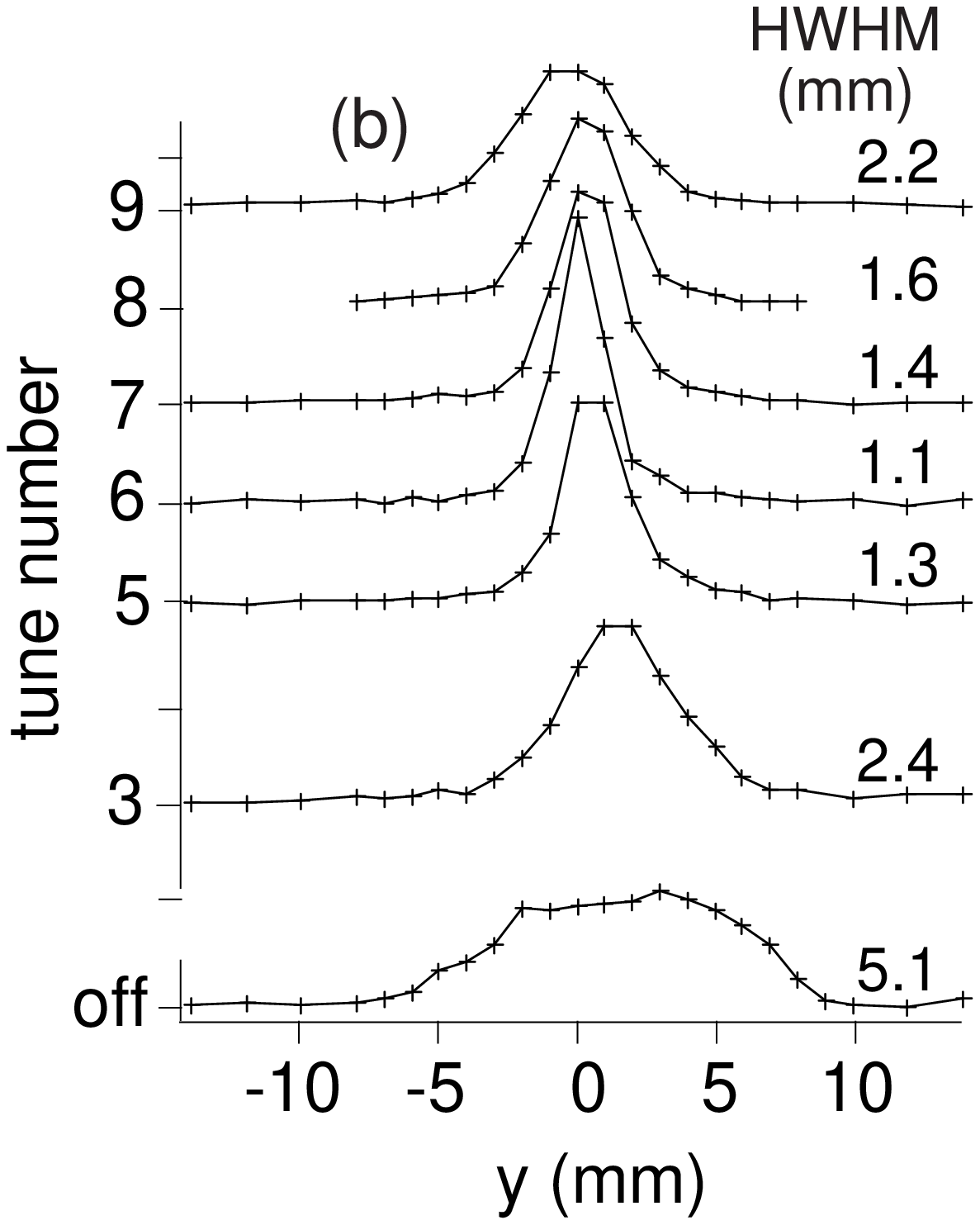}
\caption{\label{scans}
Laser probe scans (a) in $x$ and (b) in $y$ of the cesium fountain as a function of tune number in Fig~\ref{tune}. The number to the right of each peak is the half-width at half maximum (HWHM). The anomolous shoulder on the right side of some of the $x$-direction scans is caused by a reflection of the angled probe beam from the inside of an uncoated exit window. The HWHM widths are corrected for the reflection.}
\end{figure}
%
%
\begin{figure}
\includegraphics [scale = 1] {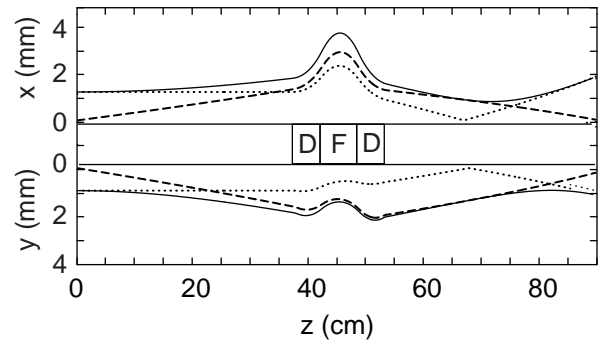}
\caption{\label{tuneb6}
HWHM beam envelope and principal rays modeled for tune 6 which focuses the atoms just upstream of the laser probe with a magnification of one. The upper (lower) plot shows the trajectories in the $x$ ($y$) direction and the focusing triplet is shown schematically inbetween. The broken lines are the calculated trajectories of an atom with zero displacement from the $z$ axis and maximum initial transverse velocities, $v_{x_i}$ = 15 mm/s and $v_{y_i}$ = 18.5 mm/s. The dotted lines are the calculated trajectories of an atom with zero transverse velocity and maximum spatial displacement from the $z$ axis:  $x_i$ = 1.2 mm and  $y_i$ = 0.95 mm. The solid lines show the calculated HWHM beam envelope. The effect of gravity is to bring the focus closer to the triplet, making object and image distance unequal for a 1:1 magnification.}
\end{figure}
%
\begin{figure}
\includegraphics [scale = 0.75] {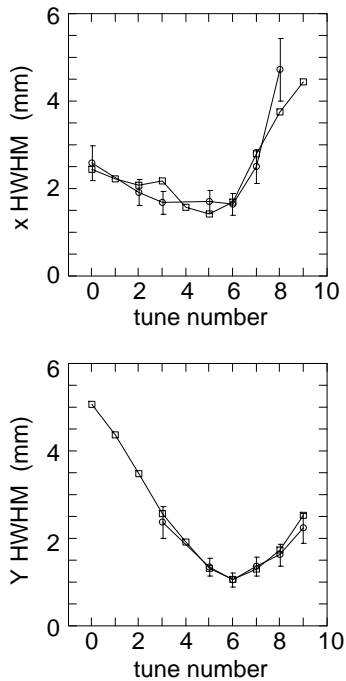}
\caption{\label{comparison}
Calculated ($\Box$) versus measured ($\odot$) beam widths (a) in $x$ and (b) in $y$ for the tunes in Fig.'s  \ref{tune},\ref{scans}.}
\end{figure}

The experimental arrangement for imaging the cesium trap, with a magnification close to one, is shown schematically in Fig.~\ref{apparatus}a. We measured the size of the cesium beam near the focus by scanning with a one mm diameter laser beam, separately in $x$ and in $y$. The laser is tuned to the cesium $6~^2S_{1/2} F = 4 \leftrightarrow 6~^2P_{3/2} F = 5$ cycling transition at 852 nm and the fluorescence light is detected by a photomultiplier tube. 

Different sets of voltages applied to the F and D lenses bring the cesium atoms to a focus at different vertical positions. The voltage pairs for the lenses that produce a minimum in both $x$ and $y$ at a given location (double waist) along the $z$ axis is called a tune and is plotted in Fig.\ref{tune}. 

The beam widths observed, at the laser probe, for the tunes of Fig\ref{tune}, are shown in Fig.\ref{scans}. In the low tune numbers, the focusing is weaker and the focus is beyond the laser probe scan, resulting in a larger size at the probe. The measured beam profiles, at the laser probe then narrow as the tune voltages increase and the focus moves to the laser probe. In tune number 6, we measured a half-width at half-maximum (HWHM) focus of 1.64 mm ($x$)  by 1.05 mm ($y$). The widths of the beam profiles, at the laser probe, then increase as the higher voltages pull the focus closer to the trap.

The calculated beam envelope (Eqs.\ref{envelope}, \ref{correction}) and principal rays for tune 6 are shown in Fig. \ref{tuneb6}. (The HWHM is related to $\sigma$ in Eq.\ref{envelope} by HWHM$ = 1.18 \sigma $.)  This focus size is consistent with the size of the trap determined from CCD camera images. 

Beam envelope calculations were performed for all of the tunes shown in Fig. \ref{tune}. In Fig.\ref{comparison}, the calculated beam widths in $x$ and $y$ at the laser probe are compared with the observed widths (from Fig.\ref{scans}). The agreement between measurement and calculation shows that the calculations can be used to make reliable predictions.

A comparison of the integrated intensity of the cesium beam profile, at the focus (tune 6 in Fig.\ref{scans}), with the cesium beam profile for no electric field (labeled off in Fig.\ref{scans}) shows a loss of about half of the cesium atoms when the beam is focused. This is the result of nonlinearities that make the linear focusing region of the triplet lens smaller than the lens aperture. If we assume that both transverse directions contribute equally to the loss, then the triplet lens has an effective aperture of 5.7 mm diameter, compared to the 8~mm center axis electrode spacing.

Some of the nonlinearity in the focusing is due to the truncation of the D-lens equipotential and can be corrected in future lens designs, but the remainder is due to inherent nonlinearities in the lens, even for a pure dipole plus sextuple field. The inherent nonlinearities can be reduced by designing the lenses with smaller $A_3$ (Eq.\ref{E}). This will result in a larger linear-focusing aperture but lower focusing strength. The lower focusing strength (Eq.\ref{correction}) can be compensated for  by using a longer lens or a stronger field. (If the lens were constructed from cylindrical electrodes, the linear region would be further reduced and the effective aperture much smaller\cite{kalnins02})

%
\begin{figure}
\includegraphics [scale = 1] {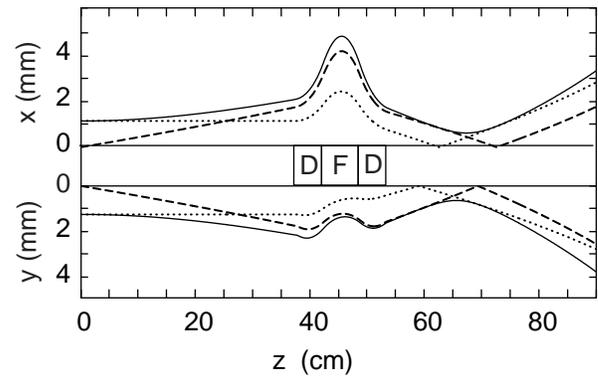}
\caption{\label{tunec9}
Modeling of the cesium fountain beam envelope and principal rays (HWHM) for focusing the trap into an image with a magnification of 0.5 (image one half the object size in both $x$ and $y$).}
\end{figure}
%
%
\begin{figure}
\includegraphics [scale = 0.75] {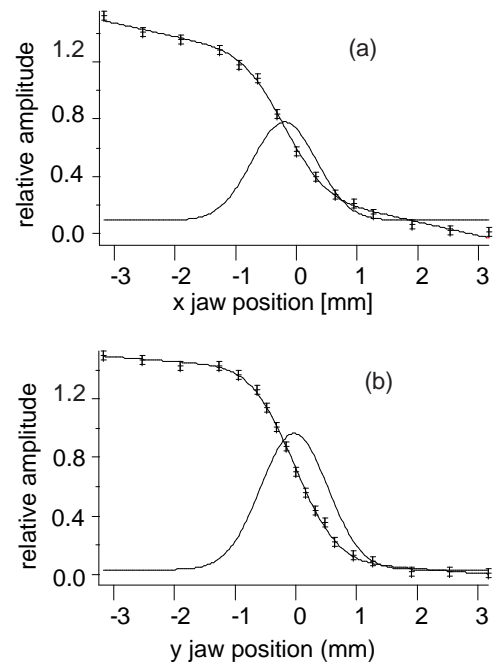}
\caption{\label{tunehalf}
Measured beam profiles (a) in $x$ and (b) in $y$ for focusing to a magnification of 0.5. The data points are the integrated beam at the detector as a function of how far the jaw is translated. The curves below  show the cesium atom intensity versus jaw position.
}
\end{figure}
%

To image the trap with a magnification of 0.5 we increased the voltages on the triplet lens to the values shown in tune 9 of Fig.\ref{tune}. Figure\ref{tunec9} shows the modeling for this tune. The voltage increase causes the focus to move closer to the lenses and away from the laser probe. The new focal point coincides with a set of four movable jaws (Fig.\ref{apparatus}b) and these were used to profile the beam (Fig.\ref{tunehalf}). The minimum waist for this tune was found to be 0.62~mm$\times$0.65~mm HWHM (compared to 1.64~mm$\times$1.05~mm for the magnification of one), and again in agreement with calculation (Fig.\ref{tunec9}).
\begin{figure}
\includegraphics [scale = 1] {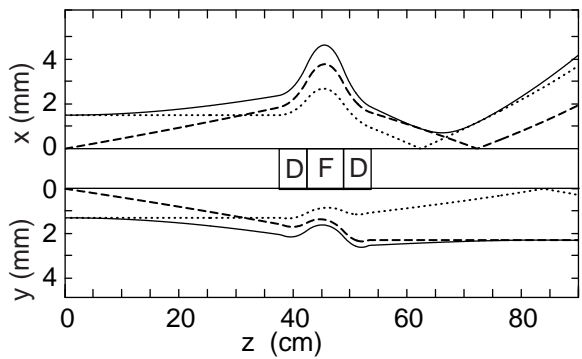}
\caption{\label{tunec10}
Modeling of the cesium fountain beam envelope (HWHM) and principal rays for focusing the cesium atoms to a 0.8 mm HWHM waist in the $x$ direction at the jaw and a nearly parallel beam of about 2.5 mm in the $y$ direction.}
\end{figure}
%
%
\begin{figure}
\includegraphics [scale = 0.80] {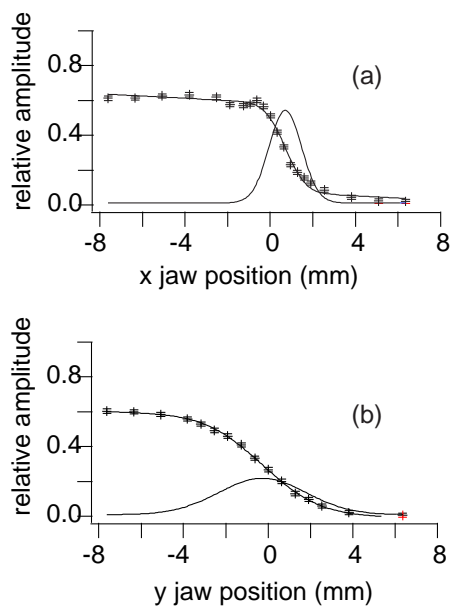}
\caption{\label{tune10data}
Cesium atom profile (a) in $x$ and (b) in $y$ obtained by scanning the jaws with the lens voltages adjusted to give a focus in the $x$ direction and a nearly parallel beam in the $y$ direction as modeled in Fig.\ref{tunec10}. }
\end{figure}

In the third set of measurements, we created a strong astigmatism in the triplet lens. In the first astigmatic  tune we produced, in the $y$ direction, a 2.5 mm HWHM beam with a low divergence, while focusing in the $x$ direction to a 0.8 mm waist at the collimation jaws. The beam envelope for this tune is shown in Fig.\ref{tunec10} and the measured profiles are shown in Fig.\ref{tune10data}. The $x$ waist is close to that of the $x$ waist of the 0.5 magnification tune (Figs.\ref{tunec9},\ref{tunehalf}) but the $y$ direction is now collimated. 

In the second astigmatic tune, the focusing in $x$ and $y$ were reversed:  The cesium fountain had the low-divergence  2.5~mm HWHM in the $x$ direction while focusing to 0.70~mm HWHM in the $y$ direction. A tune with low divergence in both the $x$ and $y$ directions could be used with a collimator could reduce the cesium-cesium scattering in a cesium fountain clock.
\section{Conclusions}
A three-element alternating-gradient electrostatic lens composed only of sextupole and dipole equipotentials and with the main fields all pointing in the same direction has been designed, built, and tested. The triplet predictably and efficiently focuses a fountain of cesium atoms to images of different magnifications or into beams with very small divergence, in agreement with calculation.
Lenses of this design have application in atomic clocks, EDM experiments, and the focusing of polar molecules in strong-field-seeking states.
\section*{Acknowledgments}
We thank Glen Lambertson and Hiroshi Nishimura for their help in reaching an understanding of the focusing of neutral atoms and molecules. We gratefully acknowledge support from the NASA Office of Biological and Physical Research, from a NIST Precision Measurements Grant, and from the Office of Science, Office of Basic Energy Sciences of the U.S. Department of Energy (DOE). The Lawrence Berkeley National Laboratory is operated for the U.S. DOE under Contract  No. DE-AC02-05CH11231.

\bibliography{focusbib}
\end{document}